\documentclass{aastex}          
\usepackage{spr-astr-addons}    

\begin{document}
%
\title{Analytical solutions for weak black hole kicks}

\shorttitle{Weak black hole kicks}
\shortauthors{M.Lingam}

\author{Manasvi Lingam\altaffilmark{1}} 
\affil{Institute for Fusion Studies, The University of Texas, Austin, TX 78712, USA}
\email{manasvi@physics.utexas.edu} 

\begin{abstract}
The black hole is modeled by a combined gravitational potential of the bulge, disk and halo and is subjected to an initial weak kick. The resulting differential equations are set up, and shown to possess analytical solutions. The effects of black hole accretion and dynamical friction are also incorporated into this analytical framework. The resultant frequencies and amplitudes are computed, and are compared with the ones obtained from numerical simulations. Within the valid range of parameters of the analytical model, the two sets of results are shown to be in reasonable agreement. It is shown that this model reproduces the linear dependence of the amplitude on the initial kick velocity, and the constant of proportionality is close to that obtained from the simulations. The analytical treatment presented is quite general, and its applications to other areas are also indicated.
\end{abstract}


\section{Introduction}
In the last decade, the potential implications of black hole mergers and subsequent recoil have been widely studied. From the results of extensive simulations using general relativity, it is now believed that the recoil velocities can be extremely large, up to a few thousands of kilometers per second \citep{cam07a,cam07b,hea09,zlo11}. Such velocities are more than sufficient to expel the black hole from the galaxy, but it has been suggested that the characteristic kick velocities tend to be much smaller, typically $\lesssim 200$ km/s, as a result of gas accretion and other associated effects \citep{bog07}. For kicks that lie in this range, one of the primary effects observed is the oscillation of the merged black hole \citep{ble08,kor08}. 

The effects of black hole mergers on galactic morphology and dynamics are expected to be significant. Several studies in this area have been undertaken, both in the context of theoretical modeling as well as predicting their observational signatures. On the theoretical front, there have been investigations of the post-Newtonian corrections to gravitational recoil resulting from black hole binaries \citep{bla05}, effects on active galactic nuclei \citep{kom08a,ble11}, hydrodynamical response \citep{cor10}, ring formation \citep{lov10a}, etc.  On the other hand, there has been plenty of ongoing analysis in understanding the observational imprints that these events produce in their aftermath. Investigations in this arena include the study of recoil effects in gaseous environments \citep{loe07,lip08,dev09,gue11,sij11}, electromagnetic signatures \citep{kom12,sch13}, black hole demographics \citep{mad04}, stellar systems in the vicinity \citep{mer09,lea09}, cusp--core conversion \citep{mer04}, offset active galactic nuclei \citep{vol08}, tidal disruption flares \citep{sto12}, etc.

Some of the most significant efforts in this area have been centered around the analysis of potential supermassive black hole candidates undergoing recoil. \cite{Kom08b} suugested that the quasar SDSS J0927+2943 was one such candidate, and several authors investigated the consequences in greater detail, while others proposed alternative theories \citep{dot09,bog09}. Subsequently, other recoiling supermassive black hole candidates were explored in the literature, such as CID-42 \citep{civ12,ble13}, the supermassive black hole in M87 \citep{bat10} and CXO J122518.6+144545 \citep{jon10}.

In this paper, we shall present an analytical model which describes the recoil of black holes when subjected to ``weak'' gravitational kicks, i.e. when the black hole oscillates with an amplitude $\lesssim$ 100 parsecs. Although such kicks are not predicted to be easily detectable \citep{ble08}, there are several advantages to the analytical approach. It permits us to highlight the similarities between the analytical model (and the divergences) and the extensive numerical simulations. Secondly, the formalism developed in this paper is quite general, and can be applied to other massive objects subjected to small perturbations \citep{cha02}. We shall focus primarily on making use of the models and results from \cite{fuj08,kor08,ble08,fuj09}, which we shall henceforth refer to as F08, KL08, BL08 and F09 respectively. 

The outline of the paper is as follows. In Section \ref{SecII}, we describe the basic assumptions behind the model, and present the simplest possible scenario with the appropriate equations of motion. In Section \ref{SecIII}, we introduce a more realistic description of the kick, and how the displacements are subsequently altered. In Section \ref{SecIV}, we introduce a new effect - Eddington mass accretion - and explore its consequences. In Section \ref{SecV}, we discuss the consequence of our analytical models, and discuss how they stack up against the numerical results. In Section \ref{SecV02}, we study the effects of constant and Bondi-Hoyle mass accretion on the dynamics of black hole kicks. Finally, in Section \ref{SecVI}, we summarize the results and explore the prospects for future work.

\section{Building the model}
\label{SecII}
In this section, we begin building our model by describing the various components that constitute it. The approach we follow is similar to that of F08, KL08 and F09. The gravitational potential arises from three distinct components, 
\begin{equation} \label{Pottot}
\Phi = \Phi_b + \Phi_g + \Phi_h,
\end{equation}
where the subscripts $b$, $g$ and $h$ indicate that these are the contributions arising from the bulge, the gaseous disk and the dark matter halo respectively. We shall assume that the bulge is modelled by the Plummer potential \citep{plu11}, the disk by the Miyamoto-Nagai potential \citep{miy75}, and the dark matter halo by the Binney logarithmic potential \citep{bin81}. This yields
\begin{eqnarray} \label{PotEqns}
\Phi_b &=& -\frac{GM_b}{\sqrt{r_b^2+R^2+z^2}}, \\ \nonumber
\Phi_g &=& -\frac{GM_g}{\sqrt{R^2+\left(a+\sqrt{z^2+b^2}\right)^2}}, \\ \nonumber
\Phi_h &=& \frac{1}{2}v_h^2 \ln \left[R^2 + \left(\frac{z}{q}\right)^2+r_c^2 \right].
\end{eqnarray}
In the above expressions, we note that $R^2 = x^2+y^2$, and that all three potentials are axisymmetric. The characteristic values associated with each of the above potentials is listed in Table \ref{Tab1}. The values for the bulge and the dark matter halo are chosen to be identical to KL08, but the spherical symmetry of the halo in KL08 has been replaced with a slight degree of asymmetry. The values for the disk are reproduced from F08 and F09. 

\begin{table}
\caption{List of parameters}
\label{Tab1} 
\begin{tabular}{|c|c|c|}
\hline 
Model & Parameter \#1 & Parameter \#2\tabularnewline
\hline 
\hline 
Bulge & $M_{b}=10^{10}M_{\odot}$ & $r_{b}=1\,\mathrm{kpc}$\tabularnewline
\hline 
Disk & $M_{g}=10^{11}M_{\odot}$ & $a=6.5\,\mathrm{kpc},$ $b=0.26\,\mathrm{kpc}$\tabularnewline
\hline 
Halo & $v_{h}=250\,\mathrm{km/s}$ & $r_{c}=2\,\mathrm{kpc}$, $q=0.9$\tabularnewline
\hline 
\end{tabular}
\end{table}

Since we are interested in developing an analytical model to describe weak kicks, we shall assume that the deviations experienced by the black hole are small, i.e. we assume that $R/L \ll 1$ and $z/L \ll 1$ where $L=\mathrm{min}\left(r_b,a,r_c\right)$. It is evident from Table \ref{Tab1} that all these length scales are of the order of a few kiloparsecs. Hence, this model is restricted to the study of kicks that result in the black hole traversing distances smaller than $100\,\mathrm{pc}$. With such an assumption, it is valid to express (\ref{PotEqns}) as follows
\begin{eqnarray}
\Phi_b &\approx& -\frac{GM_b}{r_b} \left(1-\frac{R^2}{2r_b^2}-\frac{z^2}{2r_b^2}\right), \\ \nonumber
\Phi_g &\approx& -\frac{GM_g}{a+b} \left(1-\frac{R^2}{2\left(a+b\right)^2}-\frac{z^2}{2b\left(a+b\right)}\right), \\ \nonumber
\Phi_h &\approx&  v_h^2 \ln{r_c} + \frac{v_h^2}{2 r_c^2} R^2 + \frac{v_h^2}{2 q^2 r_c^2} z^2.
\end{eqnarray}
We introduce the new variables
\begin{eqnarray}
&&\Omega_b^2 = \frac{GM_b}{r_b^3}, \quad \Omega_g^2 = \frac{GM_g}{\left(a+b\right)^3}, \quad \Omega_h^2 =  \frac{v_h^2}{r_c^2}, \\ \nonumber
&&\zeta^2 = 1+ \frac{a}{b}, \quad \Omega_{gz} = \zeta \Omega_g, \quad \Omega_{hz} = \frac{\Omega_h}{q}, \\ \nonumber
&& \Omega^2 = \Omega_b^2 + \Omega_g^2 + \Omega_h^2, \quad \Omega_z^2 = \Omega_b^2 + \Omega_{gz}^2 + \Omega_{hz}^2.
\end{eqnarray}
Using the definition of $\Phi$ from (\ref{Pottot}) and the above relations, we find that
\begin{equation} \label{Phigrad}
\frac{\partial \Phi}{\partial x} = \Omega^2 x, \quad \frac{\partial \Phi}{\partial y} = \Omega^2 y, \quad \frac{\partial \Phi}{\partial z} = \Omega_z^2 z.
\end{equation}
We shall not consider the effects of dynamical friction at this moment, since the dynamical friction damping timescale is expected to be much longer, ensuring that it does not play a significant role \citep{kor08,ble08,fuj08,lov10a}. We can solve for the displacements since we know that ${\bf{a}} = - \nabla \Phi$, which we have determined in (\ref{Phigrad}). However, there is one other factor that we need to take into account - the initial ``kick'' adminstered to the black hole. If we assume that the kick is sharp, we can model this additional force by approximating it as a delta function (in time). Hence, the additional force (per unit mass) is modelled via 
\begin{equation} \label{akick1}
{\bf{a}}_{kick} = {{\bf{v}}_0} \delta(t),
\end{equation}
where the vector $\bf{v}_0$ can be oriented in any random direction, but the magnitude $v_0$ must equal that of the initial kick velocity. It is easy to verify, given that the delta function has units of inverse time, that the above expression is also dimensionally correct. Thus, our final set of equations are given by
\begin{equation} \label{sys1}
\ddot{l} +  \Omega^2 l = v_{0l} \delta(t), \quad \ddot{z} +  \Omega_z^2 z = v_{0z} \delta(t), 
\end{equation}
where $l=x,y$. 

\section{Analytical solutions for the black hole trajectory} \label{SecIII}
In this section, we shall solve the system of differential equations given by (\ref{sys1}) and furnish an analysis of the same. The formal solution is given by
\begin{eqnarray}
l(t) &=& C_l \sin\left(\Omega t\right) + D_l \cos\left(\Omega t\right) \\ \nonumber
&+& \frac{v_{0l}}{\Omega} \int_0^t \delta(t-\tau) \sin\left(\Omega \tau \right) d\tau,
\end{eqnarray}
\begin{eqnarray}
z(t) &=& C_z \sin\left(\Omega_z t\right) + D_z \cos\left(\Omega_z t\right) \\ \nonumber
&+& \frac{v_{0z}}{\Omega_z} \int_0^t \delta(t-\tau) \sin\left(\Omega_z \tau \right) d\tau.
\end{eqnarray}
We use the boundary conditions $\dot{l}(0)=v_{0l}$, $\dot{z}(0)=v_{0z}$ and $l(0)=z(0)=0$ to simplify the above expressions. As a result, we obtain the remarkably simple expressions
\begin{equation} \label{deltasol1}
R(t) = \frac{v_{0R}}{\Omega} \sin\left(\Omega t\right), \quad z(t) = \frac{v_{0z}}{\Omega_z} \sin\left(\Omega_z t\right),
\end{equation}
where $v_{0R}^2 = v_{0x}^2 + v_{0y}^2$. This indicates that the solutions are oscillatory, and that their amplitude is linearly proportional to the initial velocity. If we assume that the initial velocity was solely in the $\hat{z}$-direction, we find that the black hole oscillates only along the $\hat{z}$-axis. This is because $v_{0R}=0$ for this system, and hence $R(t)=0$. On the other hand, if we assume that the kick was in the $x-y$ plane, it is found that $z(t)=0$, which implies that the black hole oscillates only in the $x-y$ plane. 
In this analysis, it was assumed that the black hole received a kick, wherein the acceleration was governed by (\ref{akick1}). However, this is somewhat unphysical, since the acceleration is infinite at $t=0$ and zero everywhere else. Hence, we can replace $\delta(t)$ with $\omega \exp(-\omega t)$, where $\omega$ is chosen to be suitably ``large''. Thus, we assume $\omega \gg \Omega$ and $\omega \gg \Omega_z$ hold true. With this choice of ansatz, we find that
\begin{eqnarray} \label{expo1NA}
R(t) &=& \frac{v_{0R}}{\Omega} \left(1+\frac{\omega^2}{\Omega^2+\omega^2}\right) \sin\left(\Omega t\right) \\ \nonumber
&+& {v_{0R}} \frac{\omega}{\Omega^2+\omega^2} \left[\exp(-\omega t) - \cos\left(\Omega t\right)\right],
\end{eqnarray}
\begin{eqnarray} \label{expo2NA}
z(t) &=& \frac{v_{0z}}{\Omega_z} \left(1+\frac{\omega^2}{\Omega_z^2+\omega^2}\right) \sin\left(\Omega_z t\right) \\ \nonumber
&+& {v_{0z}} \frac{\omega}{\Omega_z^2+\omega^2} \left[\exp(-\omega t) - \cos\left(\Omega_z t\right)\right].
\end{eqnarray}
It is clear that these expressions satisfy the boundary conditions specified earlier. From the above expressions, we see that the values of $R$ and $z$ are linearly proportional to $v_{0R}$ and $v_{0z}$ respectively, which matches with the results obtained in KL. As a result, if $v_{0R}=0$, then there are no oscillations along $R$ and the motion becomes purely oscillatory about the $\hat{z}$-axis. The converse holds true if $v_{0z}=0$, leading to oscillations only in the $x-y$ plane. 

The exponential term in both of the above expressions can be neglected for reasonably high values of $t$, as it results in an extremely steep falloff. In addition, we use the fact that $\omega$ is ``large'' which yields  the following expressions for $R$ and $z$.
\begin{equation} \label{Rapprox1}
R(t) = \frac{2 v_{0R}}{\Omega} \sqrt{1+\left(\frac{\Omega}{2\omega}\right)^2} \sin\left(\Omega t - \phi\right),
\end{equation}
\begin{equation} \label{zapprox1}
z(t) = \frac{2 v_{0z}}{\Omega_z} \sqrt{1+\left(\frac{\Omega_z}{2\omega}\right)^2} \sin\left(\Omega_z t - \phi_z\right),
\end{equation}
where $\tan\phi = \frac{\Omega}{2\omega}$ and $\tan\phi_z = \frac{\Omega_z}{2\omega}$. It must be emphasized however that (\ref{Rapprox1}) and (\ref{zapprox1}) are only approximate expressions, despite their simple functional form.

\section{The effects of black hole accretion} \label{SecIV}
Until now, we have implicitly proceeded with the assumption that the black hole did not accrete from its surroundings. Hence, we were able to compute the acceleration by determining the force per unit mass. In a more general scenario however, one must note that
\begin{equation}
{\bf{F}} = \frac{d{\bf{p}}}{dt} = \frac{d\left(M{\bf{v}}\right)}{dt},
\end{equation}
and hence we obtain a term involving $\dot{M}$ as well, where $M$ is the mass of the black hole. By modifying (\ref{sys1}) to incorporate the effects of mass accretion, we end up with 
\begin{equation} \label{sysBHA}
\ddot{l} + \frac{\dot{M}}{M} \dot{l} + \Omega^2 l = v_{0l} \delta(t), \quad \ddot{z} + \frac{\dot{M}}{M} \dot{z} + \Omega_z^2 z = v_{0z} \delta(t), 
\end{equation}
where $l=x,y$. It remains now to choose a suitable functional form of the mass accretion rate, and solve the problem. There are three different possibilities that present themselves. The first is the case where $\dot{M} = \mathrm{const}$, the second involves the Eddington accretion rate with $\dot{M} \propto M$, and the third is the Bondi-Hoyle accretion rate. We wish to build a fairly simple heuristic model that is analytically tractable, but also captures enough of the underlying physics. As a result, we rule out the Bondi-Hoyle accretion rate, because it involves a factor of $\left(c_s^2 + v_{rel}^2\right)^{3/2}$, which contributes a non-linearity to the problem. The first and third cases are treated in Section \ref{SecV02} separately, since they are somewhat more intricate and merit a separate discussion.

We shall work in this section with the second case since the problem can still be easily solved, but we see new effects appear. We assume $\dot{M} = \gamma M$, and defer a discussion of the value of $\gamma$ until the next section. Furthermore, we shall assume, for the sake of simplicity, that the black hole can accrete from an infinite reservoir, i.e. we shall assume that the masses $M_b$, $M_g$ and $M_h$ remain constant in time.  

There are multiple scenarios that present themselves, depending on the sign of the discriminant $\Delta = \gamma^2 - 4\Omega^2$. If $\Delta > 0$, it implies that there exist two negative real roots. Neither of these lead to the oscillatory behaviour observed in numerical simulations of KL08 and BL08. The case with $\Delta = 0$ leads to a special class of solutions, but this is a very narrow constraint on $\Delta$, and we shall return to the case later. For now, we focus on the case with $\Delta < 0$, and define $\Gamma = \frac{1}{2} \sqrt{-\Delta}$. Similarly, one can define $\Delta_z$ and $\Gamma_z$ where we replace $\Omega$ with $\Omega_z$ instead. The solutions are found to be
\begin{eqnarray}
l(t) &=& \exp\left({-\frac{\gamma t}{2}}\right) \left[C_l \sin\left(\Gamma t\right) + D_l \cos\left(\Gamma t\right)\right] \\ \nonumber
&+& \frac{v_{0l}}{\Gamma} \int_0^t \delta(t-\tau) \exp\left({-\frac{\gamma \tau}{2}}\right) \sin\left(\Gamma \tau \right) d\tau,
\end{eqnarray}
\begin{eqnarray}
z(t) &=& \exp\left({-\frac{\gamma t}{2}}\right) \left[C_z \sin\left(\Gamma_z t\right) + D_z \cos\left(\Gamma_z t\right)\right] \\ \nonumber
&+& \frac{v_{0z}}{\Gamma_z} \int_0^t \delta(t-\tau) \exp\left({-\frac{\gamma \tau}{2}}\right) \sin\left(\Gamma_z \tau \right) d\tau.
\end{eqnarray}
After the use of the boundary conditions, we find that 
\begin{equation} \label{deltasol21}
R(t) = \frac{v_{0R}}{\Gamma} \exp\left({-\frac{\gamma t}{2}}\right) \sin\left(\Gamma t\right),
\end{equation}
\begin{equation} \label{deltasol22}
z(t) = \frac{v_{0z}}{\Gamma_z} \exp\left({-\frac{\gamma t}{2}}\right) \sin\left(\Gamma_z t\right).
\end{equation}
We find that the above two expressions are quite similar to (\ref{deltasol1}), except for two crucial differences. In (\ref{deltasol1}), the solutions are purely oscillatory, while our new solutions also damp out with time. Secondly, the frequency of oscillations in (\ref{deltasol1}) is $\Omega$ ($\Omega_z$) while it is $\Gamma$ ($\Gamma_z$) for the models that undergo accretion. From the expression for $\Gamma$ ($\Gamma_z$), it can be easily verified it is smaller than $\Omega$ ($\Omega_z$). As a result, the oscillatory time scales for (\ref{deltasol21}) and (\ref{deltasol22}) are longer than their counterparts in (\ref{deltasol1}). It is also seen from (\ref{deltasol21}) and (\ref{deltasol22}) that the amplitudes are directly proportional to the velocity of the initial kick, which is consistent with the results derived earlier. Lastly, note that the limit $\gamma \rightarrow 0$ allows us to recover the expressions obtained in (\ref{deltasol1}). 
Next, we consider the ansatz where the $\delta$-function is replaced with the more realistic case of a rapidly decaying exponential, following the approach outlined in Section \ref{SecIII}. Upon carrying out the same procedure, we find that
\begin{eqnarray} \label{expofall21}
R(t) &=& \frac{v_{0R}}{\Gamma} e^{-\frac{\gamma t}{2}} \left[1+\frac{2\omega(2\omega-\gamma)}{4\Gamma^2+(2\omega-\gamma)^2}\right] \sin\left(\Gamma t\right) \\ \nonumber
&+&\frac{4\omega v_{0R}}{4\Gamma^2+(2\omega-\gamma)^2}\left[e^{-\omega t} - e^{-\frac{\gamma t}{2}} \cos\left(\Gamma t\right)\right],
\end{eqnarray}
\begin{eqnarray} \label{expofall22}
z(t) &=& \frac{v_{0z}}{\Gamma_z} e^{-\frac{\gamma t}{2}} \left[1+\frac{2\omega(2\omega-\gamma)}{4\Gamma_z^2+(2\omega-\gamma)^2}\right] \sin\left(\Gamma_z t\right) \\ \nonumber
&+&\frac{4 \omega v_{0z}}{4\Gamma_z^2+(2\omega-\gamma)^2}\left[e^{-\omega t} - e^{-\frac{\gamma t}{2}} \cos\left(\Gamma_z t\right)\right],
\end{eqnarray}
Note that these two expressions reduce to (\ref{expo1NA}) and (\ref{expo2NA}) under the limit $\gamma \rightarrow 0$. The analysis undertaken for (\ref{expo1NA}) and (\ref{expo2NA}) is also valid here, provided one accounts for the additional features included as a result of black hole accretion. These involve the existence of damping and the modification of the frequencies $\Omega$ and $\Omega_z$ to $\Gamma$ and $\Gamma_z$ respectively.
There are three different frequency scales present, and if we assume that $\omega \gg \gamma$, $\omega \gg \Gamma$ and $\omega \gg \Gamma_z$, we find that we recover (\ref{Rapprox1}) and (\ref{zapprox1}), if we replace the $\Omega$'s by their equivalent $\Gamma$'s. In addition, there is also the damping factor present, which ensures that $R(t)$ and $z(t)$ tend to zero as $t \rightarrow \infty$. 

Finally, we consider the unique case where $\Delta = 0$, and hence the relations $\gamma = 2 \Omega$ and $\gamma_z = 2\Omega_z$ hold true. We shall investigate the behaviour when the initial kick is described by the $\delta$-function, and not the decaying exponential, since the former yields a simpler and more intuitive result. The homogeneous solution can be expressed in the form $K_1 \exp(-bt)+ K_2 t \exp(-bt)$, and neither of these exhibit any oscillatory behaviour. As a result, we do not consider this case further since it does match with the results from numerical simulations.

\section{Discussion} \label{SecV}

\begin{figure*}
$$
\begin{array}{ccc}
 \includegraphics[width=5.28cm]{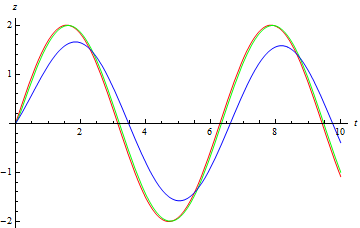} & \includegraphics[width=5.28cm]{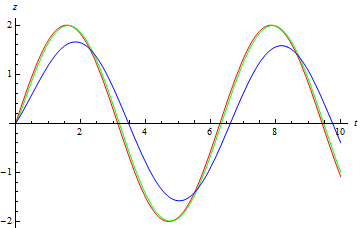} &   \includegraphics[width=5.28cm]{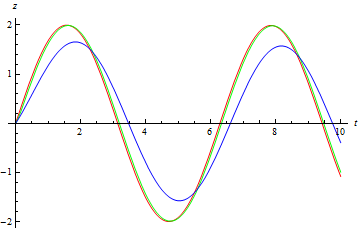}\\
 \quad\quad(a) & \quad\quad(b) &  \quad\quad(c)\\
\includegraphics[width=5.28cm]{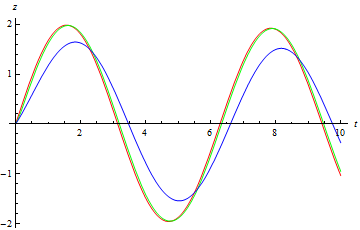} & \includegraphics[width=5.28cm]{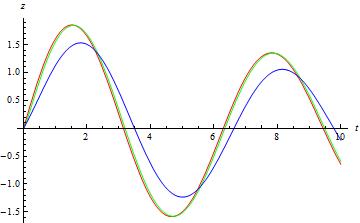} & \includegraphics[width=5.28cm]{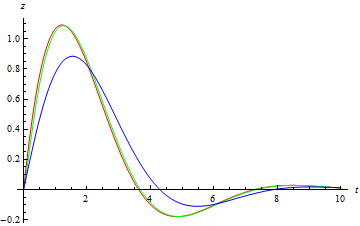}\\
 \quad\quad(d) & \quad\quad(e) & \quad\quad(f) \\
\end{array}
$$
\caption{ All of the above panels depict the plots of $z(t)$ vs $t$. In each panel, the red, green and blue curves correspond to the values $\omega=10^9\,\Omega_z$, $\omega=10\,\Omega_z$ and $\omega=\,\Omega_z$ respectively. Figs (a), (b), (c), (d), (e) and (f) depict $\gamma=0$, $\gamma = 10^{-4}\,\Omega_z$, $\gamma = 10^{-3}\,\Omega_z$, $\gamma = 10^{-2}\,\Omega_z$, $\gamma = 10^{-1}\,\Omega_z$ and $\gamma = \,\Omega$ respectively. Note that $t$ is plotted in units of $\Omega_z^{-1}$ and $z$ is plotted in units of $v_{0z}/\Omega_z$. }
\label{fig1}
\end{figure*}

Until now, our discussion was hitherto restricted to studying several classes of solutions analytically. In this section, we shall provide some simple estimates of the parameters used in our equations. In Section \ref{SecII}, it was pointed out that choosing values of $R$ and $z$ beyond 100 parsecs would violate the assumption that $R^2/L^2 \ll 1$ and $z^2/L^2 \ll 1$. We shall use this fact in determining the maximum possible value of $v_0$, for which our model remains valid. 

Let us consider the simplest scenario, described by (\ref{deltasol1}). The value of the $\Omega$'s can be calculated from Table \ref{Tab1}. The amplitude in the plane of the disk is given by $v_0/\Omega$, indicating that it has a linear dependence on the velocity. Upon substituting the appropriate values, we find that it is equal to $4.08\times10^{-3} \left(v_0\right)_{km/s}\,\mathrm{kpc}$, where $\left(v_0\right)_{km/s}$ denotes the initial kick velocity in units of $km/s$. KL08 showed that the relation between the amplitude and the velocity was linear, with a slope of $\left(9.30\pm0.53\right)\times10^{-3}$ in the above choice of units. The result that we derived is smaller than the one obtained by KL08 roughly by a factor of $2$. This arises from the differences in the parameters of the two models as well as the simplifying assumptions used in the analytical case. As a result of the simplifications employed, our model does not give rise to a $y$-intercept in the amplitude vs velocity plot. Similarly, the amplitude, normal to the plane, in the $\hat{z}$-direction can be found via $v_0/\Omega_z$ and is equal to $3.20\times10^{-3} \left(v_0\right)_{km/s}\,\mathrm{kpc}$. 

Next, we investigate the limits of our model. Since the distance travelled must be smaller than 100 parsecs, we find that the maximum possible value of the velocity is $24.5$ km/s if one considers kicks in the plane, and is $31.25$ km/s for kicks normal to the plane. As a result, this explains the discrepancy in the frequency dependence on $v_0$ for our model and KL08. The latter find that the frequency is a (weakly) linear function of the initial in-plane kick velocity. However, for the range of velocities where our model is applicable, we find that the weak linear dependence obtained by KL08 becomes insignificant. Hence, in this range, the frequency can be treated as being nearly independent of the velocity, which is also consistent with our results. However, our model yields a (velocity-independent) frequency that is roughly an order of magnitude higher than that obtained by KL08.

Next, we consider the scenario where black hole accretion plays a significant role. The simplest scenario is where the kick is modelled via a $\delta$-function, and the governing equations are (\ref{deltasol21}) and (\ref{deltasol22}). In order to proceed further, we need to determine the value of $\gamma$. We shall use the following expression from \cite{koc13}
\begin{equation}
\gamma^{-1} = t_{\mathrm{Edd}} = 4\times 10^7 \left(\frac{\epsilon/(1-\epsilon)}{0.1}\right)\left(\frac{L}{L_{\mathrm{Edd}}}\right)^{-1}\,\mathrm{yr},
\end{equation}
where $\epsilon$ is the radiative efficiency, $L$ is the luminosity and $L_{\mathrm{Edd}}$ is the Eddington luminosity. If we assume that $\epsilon \sim 0.1$ and $L \sim L_{\mathrm{Edd}}$, we find that $\gamma \approx 7.93 \times 10^{-16} s^{-1} \approx 0.1 \Omega$. As a result, one finds that $\Gamma \approx \Omega$, and $\Gamma_z \approx \Omega_z$. Hence, the amplitudes still remain linear in the velocity, and are virtually identical to the results discussed above, except that they are slightly higher. 

Lastly, we consider the effects of dynamical friction. We will turn our attention to the simplest case, where there is no black-hole accretion taking place. The stellar bulge exerts a dynamical friction which can be easily estimated via Chandrasekhar's dynamical friction formula \citep{bin87} which states that the drag experienced is given by
\begin{equation}
{\bf{F}} = - \beta {\bf{v}},
\end{equation}
where $\bf{v}$ is the velocity of the black hole, and $\beta$ is given by
\begin{equation} \label{betaexact}
\beta = 16\pi^2 \ln{\Lambda} G^2 M (M+m) \frac{\int_0^v f(r,u) u^2 du}{v^3},
\end{equation}
where $M$ is the mass of the black hole, $m$ represents the mass of each individual black hole, $\ln{\Lambda}$ is the Coulomb logarithm, and $f$ is the distribution function of the stellar bulge. From Section \ref{SecII}, we know that the bulge is modelled by the Plummer potential--density pair, which has a well-known polytropic distribution function. In addition, we have shown that our model is only valid for initial kicks whose \emph{maximum} velocities are in the range $20$-$30$ km/s. The escape velocity $v_{esc}$ of the Plummer model is known to be $\sqrt{-2\Phi}$ which can be approximated by $\sqrt{2GM_b/r_b}$ since the kicks result only in small perturbations. The escape velocity is approximately $300$ km/s, and hence the approximation $v/v_{esc} \ll 1$ is (mostly) a valid one. With this approximation and using the fact that $r/r_b \ll 1$, we can simplify (\ref{betaexact}) to yield
\begin{equation} \label{betaapprox}
\beta = \frac{128\sqrt{2}}{7\pi}\ln{\Lambda} \sqrt{\frac{G}{M_b r_b^3}} M^2,
\end{equation}
where we also used $M \gg m$ and the explicit expression for the Plummer distribution function \citep{bin87,lin14}. Since $\beta$ is independent of $v$, we find that the dynamical friction is linearly proportional to the velocity. As we are considering the case with no black hole accretion, we are interesting in calculating the force per unit mass. Combining (\ref{betaapprox}) with  (\ref{sys1}), we find that the governing equations are
\begin{equation}
\ddot{l} + \zeta \dot{l} + \Omega^2 l  = v_{0l} \delta(t), \quad \ddot{z} + \zeta \dot{z} + \Omega_z^2 z  = v_{0z} \delta(t),
\end{equation}
where $\zeta = \beta/M$. Note that the above equation is exactly the same as (\ref{sysBHA}), since we assumed that $\gamma = \dot{M}/M$ was constant in time. Hence, the case with zero black hole accretion and a non-zero dynamical friction is mathematically identical to the case with non-zero black hole accretion and zero dynamical friction. As a result, the mathematical results are identical with $\gamma$ replaced by $\zeta$ instead. Let us now estimate the value of $\zeta$. We assume $\ln{\Lambda} \approx 3$, and that $M \approx 10^8 M_\odot$ and substitute them into (\ref{betaapprox}). After simplification, we find that $\zeta = 0.01 \Omega_b \approx 6.74 \times 10^{-17} s^{-1}$. Hence, we find that $\zeta \ll \Omega$ is a valid assumption, indicating that over timescales of $\Omega^{-1}$, dynamical friction does not play a significant role in our model. This ensures that our neglect of dynamical friction was quite a reasonable assumption. For our choice of parameters and models, we find that $\zeta = 0.085 \gamma$, indicating that the black hole accretion induced damping is somewhat more significant that the one caused by dynamical friction. We have only considered the case where only one of these two effects (black hole accretion and dynamical friction) are present. It is possible, in principle, to deduce analytical solutions even when the two mechanisms are considered simultaneously. However, the resultant solutions are mathematically complex, involving hypergeometric and Laguerre functions. The solutions are not particularly illuminating either, i.e. we do not see any unexpected behaviour.  

Lastly, we note that our discussion has not hitherto considered the case where the kick is described by an exponential falloff, as opposed to the $\delta$-function. This is because of the fact that there is no \emph{a priori} means of estimating the value of $\omega$, which represents the inverse timescale for the falloff. However, it does alter the behaviour $R(t)$ and $z(t)$, evident from \ref{fig1}, which is a graphical representation of (\ref{expofall22}). This plot reduces to (\ref{expo2NA}) under the limit $\gamma \rightarrow 0$, which is depicted in the first panel of the figure. The corresponding plots for $R(t)$ are not given, since the behaviour is exactly the same, provided one replaces $v_{0z}$, $\Gamma_z$ and $\Omega_z$ with $v_{0R}$, $\Gamma$ and $\Omega$ respectively. The choice of $\omega = 10^{9} \,\Omega_z$ is the closest to representing a true ``kick'', since it implies that the exponential falloff of the kick is sharp; the timescale $\omega^{-1}$ is of the order of a few years. The other values of $\omega$ have been plotted to show that this model can also be used to model phenonema, wherein the external force is not really a ``kick'' per se, but a slower exponential falloff. It is found that the function $z(t)$ is not significantly altered by $\omega$ only when it is roughly the same order of magnitude as $\Omega_z$. In other words, for all $\omega \gg \Omega_z$, the curves nearly overlap with one another. Also note that the last panel, with the highest value of $\gamma$ exhibits the maximum amount of damping, which is to be expected. 

\section{Alternative black hole accretion rates and their consequences} \label{SecV02}
In the discussion and derivations presented after (\ref{sysBHA}), we used the Eddington accretion rate as our choice, as it yielded analytically tractable and simple solutions. In this section, we shall consider the other two possibilities, and explore their consequences. Before doing so, let us note that we used two different methods to model the kick - the first was via the delta function, and the second was by introducing an exponential falloff. The most realistic ``kick'' for the latter scenario is modelled via $\omega = 10^{9} \,\Omega_z$ in Figure \ref{fig1}, since the kick must be sharp and have a rapid exponential decay. It is found that this particular choice does not deviate much from the first case wherein the kick is modelled by a delta function. As a result, we shall consider (\ref{sysBHA}) alone, and solve it for the cases with $\dot{M} = \mathrm{const}$ and the Bondi-Hoyle accretion rate.

\subsection{Constant accretion rate}

\begin{figure}
\quad\quad\quad \includegraphics[width=5.68cm]{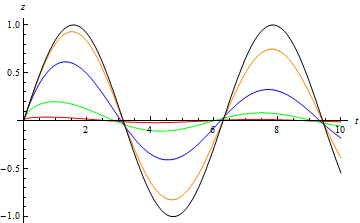} \\
\caption{The figure depicts $z(t)$ vs $t$. The red, green, blue, orange and black curves correspond to the values of $\epsilon=0.01$, $\epsilon=0.1$, $\epsilon=1$, $\epsilon=10$ and $\epsilon=100$ respectively, where $\epsilon= \Omega M_0/\alpha$. Note that $t$ is plotted in units of $\Omega_z^{-1}$ and $z$ is plotted in units of $v_{0z}/\Omega_z$.}
\label{fig2}
\end{figure}

If we assume that $\dot{M} = \alpha$, with $\alpha$ held constant, then it immediately follows that $M(t) = M_0 + \alpha t$, and $M_0$represents the black hole mass at time $t=0$. Let us consider the first equation in (\ref{sysBHA}), since the second one is found simply by replacing $\Omega$ by $\Omega_z$ and $v_{0l}$ by $v_{0z}$. Hence, our equation is given by
\begin{equation} \label{constevol}
\ddot{l} + \frac{\alpha}{\alpha t + M_0} \dot{l} + \Omega^2 l = v_{0l} \delta(t),
\end{equation}
and we shall use the same boundary conditions as in the prior sections. We recover exact solutions given by
\begin{eqnarray} \label{Rbess}
R(t)&=&\frac{v_{0R}}{\Omega}\Bigg[J_{1}\left(\frac{\Omega M_{0}}{\alpha}\right)Y_{0}\left(\frac{\Omega M_{0}}{\alpha}\right) \nonumber \\
&&-J_{0}\left(\frac{\Omega M_{0}}{\alpha}\right)Y_{1}\left(\frac{\Omega M_{0}}{\alpha}\right)\Bigg]^{-1}\times \nonumber \\
&&\Big[J_{0}\left(\frac{\Omega M_{0}}{\alpha}\right)Y_{0}\left(\frac{\Omega M_{0}}{\alpha}+\Omega t\right) \nonumber \\
&&-Y_{0}\left(\frac{\Omega M_{0}}{\alpha}\right)J_{0}\left(\frac{\Omega M_{0}}{\alpha}+\Omega t\right)\Bigg],
\end{eqnarray}
\begin{eqnarray} \label{zbess}
z(t)&=&\frac{v_{0z}}{\Omega}\Bigg[J_{1}\left(\frac{\Omega M_{0}}{\alpha}\right)Y_{0}\left(\frac{\Omega M_{0}}{\alpha}\right) \nonumber \\
&&-J_{0}\left(\frac{\Omega M_{0}}{\alpha}\right)Y_{1}\left(\frac{\Omega M_{0}}{\alpha}\right)\Bigg]^{-1}\times \nonumber \\
&&\Big[J_{0}\left(\frac{\Omega M_{0}}{\alpha}\right)Y_{0}\left(\frac{\Omega M_{0}}{\alpha}+\Omega t\right) \nonumber \\
&&-Y_{0}\left(\frac{\Omega M_{0}}{\alpha}\right)J_{0}\left(\frac{\Omega M_{0}}{\alpha}+\Omega t\right)\Bigg].
\end{eqnarray}

In Fig \ref{fig2}, we plot (\ref{zbess}) for different values of $\epsilon = \Omega M_0/\alpha$. It is seen that the plots are highly sensitive to the values of $\epsilon$, but they grow progressively sinusoidal for increasingly higher values of $\epsilon$. This can be explained in the following manner. First, note that the ratio of $\alpha/M_0$ has the units of frequency, and it can be interpreted as a measure of the growth rate, i.e. a higher value of this quantity leads to the black hole's quicker growth. In Section \ref{SecV}, the analogies between black hole accretion and dynamical friction were pointed out. In other words, a higher value of $\alpha/M_0$ leads to a higher damping rate. A higher value of $\alpha/M_0$, also leads to a lower value of $\epsilon$ as seen from the formula. Hence, we see that the curves with the lowest values of $\epsilon$ exhibit the highest damping. In the limit when $\alpha \rightarrow 0$, we see that (\ref{constevol}) reduces to (\ref{sys1}), which has been shown to possess sinusoidal solutions. As a result, this explains why higher values of $\epsilon$ lead to progressively sinusoidal curves in Fig \ref{fig2}.

\subsection{Bondi-Hoyle accretion}

\begin{figure*}
$$
\begin{array}{ccc}
 \includegraphics[width=5.28cm]{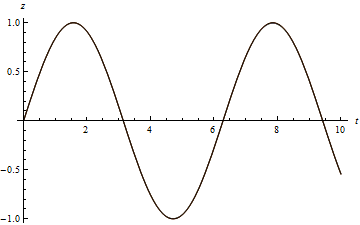} & \includegraphics[width=5.28cm]{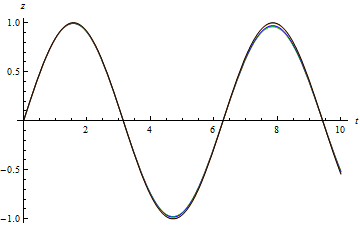} &   \includegraphics[width=5.28cm]{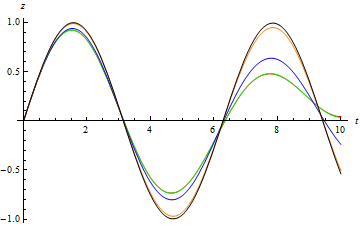}\\
 \quad\quad(a) & \quad\quad(b) &  \quad\quad(c)\\
\includegraphics[width=5.28cm]{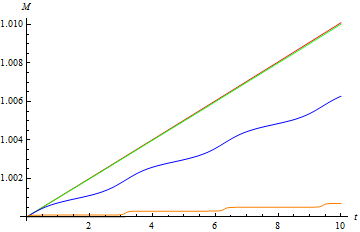} & \includegraphics[width=5.28cm]{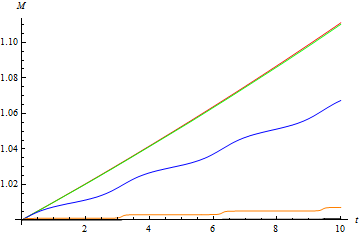} & \includegraphics[width=5.28cm]{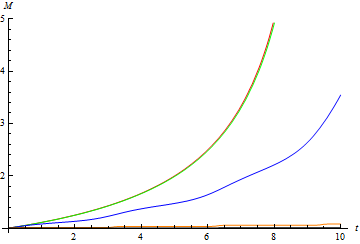}\\
 \quad\quad(d) & \quad\quad(e) & \quad\quad(f) \\
\end{array}
$$
\caption{Panels (a), (b) and (c) are plots of $z(t)$ vs $t$. Panels (d), (e) and (f) are plots of $M(t)$ vs $t$. Panels (a) and (d) hold $\nu = 10^{-3}$ fixed. Panels (b) and (e) hold $\nu = 10^{-2}$ fixed. Panels (c) and (f) hold $\nu=10^{-1}$ fixed. The red, green, blue, orange and black curves correspond to the values of $\mathcal{M}=0.01$, $\mathcal{M}=0.1$, $\mathcal{M}=1$, $\mathcal{M}=10$ and $\mathcal{M}=100$ respectively. Note that all these curves have been plotted in terms of dimensionless units.}
\label{fig3}
\end{figure*}

Let us now consider the scenario where the black hole accretes via Bondi-Hoyle accretion. The accretion rate is known to be
\begin{equation} \label{BondHoyacc}
\dot{M} = 4\pi \rho_0 \frac{G^2 M^2}{v^3_{\mathrm{rms}}},
\end{equation}
where $\rho_0$ represents the ambient medium density and $v_{\mathrm{rms}}$ denotes the gas velocity relative to that of the black hole, obeying the relation $v^2_{\mathrm{rms}} = c_s^2 + V^2$. Here, $c_s$ denotes the sound speed and $V$ denotes the relative velocity of the black hole \citep{koc13}. Before proceeding further, we note that fully analytical solutions can be obtained in the limits where  $V \ll c_s$ - the relative velocity of the black hole with respect to the surrounding medium is much smaller than the sound speed. In this event, we find that $\dot{M} = \mu M^2$, where $\mu = 4\pi \rho_0 G^2 M^2 c_s^{-3}$. The first equation of (\ref{sysBHA}) reduces to
\begin{equation} \label{BHanaly}
\ddot{l} + \left(\frac{\mu M_0}{1-\mu M_0 t}\right) \dot{l} + \Omega^2 z = v_{0l} \delta(t),
\end{equation}
and $M_0$ represents the initial mass at $t=0$. The $z$ component is found by replacing $\Omega$ and $v_{0l}$ with $\Omega_z$ and $v_{0z}$ respectively. A further analysis of the above equation is unnecessary since the similarities with (\ref{constevol}) are evident. In particular, it is seen that the second terms in (\ref{constevol}) and (\ref{BHanaly}) are exactly similar, if one identifies $\alpha$ with $\mu M_0$ and reverses the sign in the denominator of the second term in (\ref{constevol}). 

In Sections \ref{SecII} - \ref{SecV}, we have seen that the final solution for $R$ and $z$ depend on $v_{0R}$ and $v_{0z}$ respectively. Effectively, this implies that a in-plane kick excites in-plane oscillations and a perpendicular kick (in the $z$-direction) results in oscillations in the $z$-direction; in other words, the motions in the $x$-$y$ plane and in the $z$-direction are decoupled. For the sake of simplicity, we shall consider the latter scenarion. In this event, our set of three equations reduce to just one equation - the second one in (\ref{sysBHA}). In this scenario, we find that the $z$-component couples to (\ref{BondHoyacc}) to yield two interlinked equations
\begin{equation} \label{BHzrate}
\dot{M} = 4\pi \rho_0 \frac{G^2 M^2}{\left(c_s^2+\dot{z}^2\right)^{3/2}},
\end{equation}
\begin{equation} \label{BHaccz}
\ddot{z} +  \frac{\dot{M}}{M} \dot{z} + \Omega_z^2 z = v_{0z} \delta(t).
\end{equation}
In order to render them more transparent and easier to solve, we scale each of these variables to make them dimensionless. Our new variables are now given by
\begin{equation} \label{dimlessvari}
\bar{t} = \Omega_z t,\quad \bar{z} = \frac{\Omega_z}{v_{0z}} z,\quad \bar{M} = \frac{1}{M_0} M,
\end{equation}
and we introduce the following two dimensionless variables
\begin{equation}
\mathcal{M} = \frac{v_{0z}}{c_s},\quad \nu = \frac{\mu M_0}{\Omega_z},
\end{equation}
which both possess an elegant physical interpretation. $\mathcal{M}$ can be interpreted as a Mach number of the initial kick velocity to that of the isothermal sound speed. $\nu$ is interpreted as ratio of the timescale of oscillation, on the order of $\Omega_z^{-1}$, and that of the timescale of accretion, on the order of $\left(\mu M_0\right)^{-1}$. We shall drop the overbars henceforth, and note that the two equations below are expressed purely in terms of \emph{dimensionless} quantities.
\begin{equation} \label{NDBHzrate}
\ddot{z} + \frac{\dot{M}}{M} \dot{z} + z = \delta(t),
\end{equation}
\begin{equation} \label{NDBHaccz}
\dot{M} = \nu M^2 \left(1+ {\mathcal{M}}^2 \dot{z}^2 \right)^{-3/2}.
\end{equation}
Note that the $\delta$ function in (\ref{NDBHzrate}) is actually a function of $\bar{t}$, and is also dimensionless. The boundary conditions for this system are given by $M(0) = 1$, $z(0) = 0$ and $\dot{z}(0) = 1$.

The resultant figures have been plotted in Fig \ref{fig3}. In (\ref{NDBHaccz}), if we set $\mathcal{M} = 0$, then we find that it decouples from (\ref{NDBHzrate}). Furthermore, it is found that the mass (in dimensionless units) grows as $M= \left(1-\nu t\right)^{-1}$ in this case. In other words, we hit a singularity by the time $t=\nu^{-1}$. If we let $\nu$ be of order unity, then $t$ blows up at around unity as well. Since our domain in Fig \ref{fig3} ranges from $0$ to $10$, we consider only those values of $\nu$ that are at least one order smaller than unity. This is also physically justified since the timescale of accretion is expected to be longer than that of oscillation, and by the interpretation of $\nu$ outlined above, this results in smaller values than unity.

There are several interesting trends that emerge from Fig \ref{fig3}. In all the cases, we see that the higher the value of $\mathcal{M}$, the slower is the black hole's mass growth. This is to be expected by mathematically inspecting (\ref{NDBHaccz}), but it has a simpler physical explanation. The Bondi radius is proportional to $v^2_{\mathrm{rms}}$, resulting in a fairly strong inverse dependence on $\mathcal{M}$. As the Bondi radius represents the region over which the black hole can accrete, a higher $\mathcal{M}$ leads to a lower Bondi radius and accretion. 

Next, let us consider the black hole trajectories depicted in Fig \ref{fig3}. As the value of $\nu$ increases, we see that the curves with different values of $\mathcal{M}$ grow increasingly distinguishable. The reason is simple: the lower the value of $\nu$, the closer is the behaviour to that of the scenario considered in Section \ref{SecIII} with no black hole accretion. As a result, the lower values of $\nu$ ensures that all the curves cluster together, undergoing minimal damping. In panel (c) of Fig \ref{fig3}, we see that curves with lower values of $\mathcal{M}$ exhibit greater damping. This is again on expected lines - the lower the value of $\mathcal{M}$, the higher is the accretion rate, and hence the greater is the damping.  

\section{Conclusion} \label{SecVI}
In general, modelling the dynamics of a black hole is a highly complex process as it entails the inclusion of multiple processes such as the gravitational potentials of the bulge, disk and halo, the presence of dynamical friction, accretion onto the black hole, etc. A fully consistent theory requires the use of extensive numerical simulations, as undertaken in KL08 and BL08 respectively. However, we have shown that a simplified description of the black hole recoil can be constructed, which can then be solved analytically. 

Our model includes the contributions from all the gravitational potentials, and even allows for simplified accretion models to be taken into account. We also illustrate that the inclusion of dynamical friction does not play a significant role, as the timescale is much longer than the period of oscillations. We also show that dynamical friction can be described by the same mathematics as black hole accretion, i.e. they both act as damping processes where the retarding force is linearly proportional to the velocity. As a result, one can either model a black hole with no accretion and non-zero dynamical friction, or vice-versa. 

In Sections \ref{SecIII} and \ref{SecIV}, we derived different classes of analytical solutions which depended on the specific ansatz chosen for the force produced by the kick. In Section \ref{SecV} the algebraic expressions were evaluated to yield explicit expressions that were compared against those obtained by KL08. It was shown that the two results were in reasonable agreement with one another for the range of displacements (and velocity kicks) where our analytical models were valid. In Section \ref{SecV02}, we generalize the results of Section \ref{SecIV} by considering the cases of constant mass accretion and Bondi-Hoyle accretion, and the latter scenario was shown to couple the mass accretion and dynamics together. In each case, we see that the overall behaviour is exactly as one would expect from basic physical reasoning, confirming that the models accurately capture the essential physics. 

It is possible to extend the formalism to include velocity anisotropy in the dynamical friction by making use of the results from \cite{bin77}. In conclusion, we note that we can adapt our methodology and use it to study associated phenomena, since the basic tenets of our model are quite general. A natural extension of this approach would involve the extension of our model to study the dynamics of supermassive black holes subjected to stochastic kicks \citep{cha02}, or in modeling the stochastic oscillations of relativistic disks \citep{har12}.

\acknowledgments
The author thanks Philip Morrison for his support and guidance, and Santiago Jose Benavides for providing a couple of clarifications. This work was supported by the U.S. Dept. of Energy Contract \# DE-FG05-80ET-53088.

\nocite{*}
\bibliographystyle{spr-mp-nameyear-cnd}
\bibliography{biblio-u1}

\end{document}